\documentclass[12pt]{article}

\usepackage[margin=1in]{geometry}

\usepackage{setspace}

\usepackage[T1]{fontenc}
\usepackage{bold-extra}

\usepackage{caption}
\usepackage{graphicx}
\usepackage{float}
\usepackage{url}

\usepackage{amsmath,amssymb}
\usepackage{xfrac}

\usepackage[ruled,vlined]{algorithm2e}
\usepackage{algpseudocode}

\usepackage{indentfirst} 

\usepackage{hyperref}
\usepackage{xcolor}
\definecolor{Brown}{RGB}{110,55,0}
\definecolor{Grey}{RGB}{225,225,225}
\definecolor{UrlColor}{RGB}{0,0,204}
\hypersetup{
    colorlinks,
    linkcolor={Brown}, 
    citecolor={Brown}, 
    urlcolor={UrlColor} 
}

\usepackage{authblk}
\usepackage[sort&compress,round]{natbib}

\newcommand{\Ne}{\(N_e\)}

\newcommand{\eg}{\emph{e.g.}}
\newcommand{\ie}{\emph{i.e.}}

\date{}
\title{\vspace{-2em}\textbf{Not Just \(\boldsymbol{N_e}\) \(\boldsymbol{N_e}\)-more: New Applications for SMC from Ecology to Phylogenies}}
\author[1,2,3$\star$]{David Peede}
\author[4]{Trevor Cousins}
\author[5, 6, 7]{Arun Durvasula}
\author[8]{Anastasia Ignatieva}
\author[9]{Toby G. L. Kovacs}
\author[10]{Alba Nieto}
\author[11]{Emily E. Puckett}
\author[1,2$\star$]{Elizabeth T. Chevy}
\affil[1]{Department of Ecology, Evolution, and Organismal Biology, Brown University, Providence, RI, USA}
\affil[2]{Center for Computational Molecular Biology, Brown University, Providence, RI, USA}
\affil[3]{Institute at Brown for Environment and Society, Brown University, Providence, RI, USA}
\affil[4]{Department of Genetics, University of Cambridge, Cambridge, UK}
\affil[5]{Division of Epidemiology, Department of Population and Public Health Sciences, Keck School of Medicine, University of Southern California, Los Angeles, CA, USA}
\affil[6]{Center For Genetic Epidemiology, Keck School of Medicine, University of Southern California, Los Angeles, CA, USA}
\affil[7]{Department of Quantitative and Computational Biology, University of Southern California, Los Angeles, CA, USA}
\affil[8]{Department of Statistics, University of Oxford, Oxford, UK}
\affil[9]{School of Life and Environmental Sciences, University of Sydney, Sydney, NSW, Australia}
\affil[10]{ISYEB, EPHE, Université PSL, Sorbonne Université, MNHN, IRD, CNRS, 75005 Paris, France}
\affil[11]{Department of Biological Sciences, University of Memphis, Memphis, TN, USA}

\begin{document}
\maketitle

\begin{abstract}
    Genomes contain the mutational footprint of an organism's evolutionary history, shaped by diverse forces including ecological factors, selective pressures, and life history traits. The sequentially Markovian coalescent (SMC) is a versatile and tractable model for the genetic genealogy of a sample of genomes, which captures this shared history. Methods that utilize the SMC, such as \texttt{PSMC} and \texttt{MSMC}, have been widely used in evolution and ecology to infer demographic histories. However, these methods ignore common biological features, such as gene flow events and structural variation. Recently, there have been several advancements that widen the applicability of SMC-based methods: inclusion of an isolation with migration model, integration with the multi-species coalescent, incorporation of ecological variables (such as selfing and dormancy), inference of dispersal rates, and many computational advances in applying these models to data. We give an overview of the SMC model and its various recent extensions, discuss examples of biological discoveries through SMC-based inference, and comment on the assumptions, benefits and drawbacks of various methods.
\end{abstract}

\noindent $\star$ Corresponding authors: \\
\noindent \url{david_peede@brown.edu} (DP) \\
\noindent \url{elizabeth_chevy@brown.edu} (ETC) 

\subsubsection*{Keywords}
\noindent sequentially Markovian coalescent, demographic inference, gene flow, conservation biology, ARG reconstruction

\subsection*{Significance statement}
Inferring  historical population size (\Ne) from genome sequences is common across fields such as population genetics, natural history, and ecology.
The methods that generate these inferences (\eg{} \texttt{PSMC}, \texttt{MSMC}) share a common statistical framework, the sequentially Markovian coalescent.
Recent extensions to this framework now allow for the inference of quantities and forces beyond \Ne, such as population structure, gene flow, or ecologically relevant life history traits, widening its applicability to larger and more varied datasets.
In an effort to extend awareness of these new possibilities across fields, we review how these extensions work, where they have been applied, and when they are (and are not) an effective choice.

\section{Introduction}
Detecting signals of historic demographic events, such as population size changes, given DNA samples from a population is a fundamental goal of population geneticists, natural historians, and conservation biologists alike. 
The coalescent \citep{kingman_coalescent_1982, Kingman1980_math_of_diversitys, Kingman1982_geaneology_of_pops} and its subsequent extensions that account for recombination \citep{hudson_properties_1983,Griffiths_two_loucs_arg_1991} provided the probabilistic foundations for the development of statistical inference methods that estimate demographic parameters from sample genealogies---for a comprehensive overview, see \citet{wakeley_coalescent_2008}.
These coalescent models trace ancestral relationships backwards in time for a set of sampled lineages, where a coalescence event occurs when two lineages merge into a single ancestral lineage at some point in the past.
A key parameter of interest is the effective population size (\Ne), which has been defined in various ways in the literature to describe quantities such as the level of genetic diversity, the strength of genetic drift, and the efficacy of selection \citep{Charlesworth2009Ne}.
An alternative yet convenient perspective is to define \Ne{} as the inverse of the coalescence rate, as this definition can flexibly encapsulate various properties of an evolutionary model. 

For non-recombining segments of DNA---such as mitochondrial DNA (mtDNA) and chloroplast DNA (cpDNA)---the genealogical history of a sample can be represented by a single tree structure. 
However, a recombination event will split a segment of DNA into two, resulting in distinct genealogical histories for the two recombinant segments. 
The coalescent with recombination (CwR) was originally formulated as a backwards in time process that models the ancestral relationships for a set of samples while accounting for recombination events \citep{hudson_properties_1983}.
The CwR can be conceptualized in two distinct ways: as a collection of local trees or as a graph structure known as the ancestral recombination graph \citep[ARG;][]{griffiths_ancestral_1996, Griffiths_two_loucs_arg_1991, Griffiths1997_an_arg}.
The ARG can encode both coalescence events (where lineages merge) and recombination events (where lineages split), thereby describing the complete evolutionary history for a set of samples at every genomic position.
In contrast to its initial formulation, the CwR can alternatively be viewed as a sequential process along the genome rather than backwards in time \citep{wiuf_recombination_1999, Wiuf1999_ancestry_with_recomb}.
In this perspective, a local genealogy persists along the genome until a recombination event induces a change in tree structure.
Inferring the latent ARG of a set of sequences is notoriously difficult, due to the observed data (\ie, the distribution of mutations) not being not very informative about the underlying genealogical history, combined with the fact that the space of possible genealogical histories is extremely large.

To overcome the computational challenges of inference under the CwR, \citet{mcvean_approximating_2005} proposed the sequentially Markovian coalescent (SMC) as a tractable approximation to this model, which was further refined by \citet{marjoram_fast_2006}.
The SMC framework ushered in a new suite of methods based on coalescent Hidden Markov Models \citep[HMMs;][]{li_inference_2011,schiffels_inferring_2014,hobolth_genomic_2007} designed to infer demographic parameters from genomic data.
Notably, \citeauthor{li_inference_2011}’s \texttt{PSMC} method leveraged the SMC in the transition matrix of its HMM to infer the \Ne{} trajectory of humans over time, using a diploid sequence of just one individual.
Although the original SMC framework follows the assumptions of the standard coalescent, recent theoretical work has extended the SMC to account for factors such as migration \citep{wang_tracking_2020}, structural variation \citep{ignatieva_length_2024}, and selfing \citep{strutt_joint_2023}.
These extensions modify the SMC’s probabilistic model or associated HMM approach to infer quantities beyond \Ne{}.

This review presents these recent extensions to the SMC that broaden its relevance to fields such as natural history and conservation biology.
After a brief primer describing the fundamentals of the SMC framework, we divide its extensions into three broad classes: those doing \emph{joint parameter inference}, those doing \emph{lineage partitioning}, and those building upon the SMC to explicitly infer sample genealogies (in the form of ARGs).
For each class, we describe how recent methods extend the SMC, then review examples of their application to empirical (or simulated) data.
Finally, as plausible uses of the SMC expand across fields we discuss how empiricists can be best served by these new approaches, and where we expect the SMC framework will continue to prove particularly useful.

\section{SMC Primer} 
\label{sec:primer}
We begin by briefly defining the SMC model, describing how the SMC is used for inference in the \texttt{PSMC} framework, and reviewing subsequent methodological advances for inferring historical population sizes and coalescence rates using SMC-based approaches.

\subsection{Coalescent theory primer}
\label{sec:primer:coal}
Kingman's coalescent \citep{kingman_coalescent_1982, Kingman1980_math_of_diversitys, Kingman1982_geaneology_of_pops} is a probabilistic model that describes the ancestral relationships at a single non-recombining locus for a set of lineages in an idealized population (that is, panmictic with constant size, discrete generations, and no selection) of \(N\) diploid individuals, when \(N\) is large.
Rescaling time (\(t\)) leads to a continuous-time limit, where time is measured in coalescent units going backwards in time (\ie, \(t = 0\) is the present).
This can be converted back to time measured in generations through multiplying by a factor of \(2N_e\), the expected time for a pair of lineages to coalesce. Note that the effective population size \Ne{} is in general not equal to the census size of the population, reflecting the fact that the accumulation of genetic diversity is affected by violations of the neutral assumptions of the model.

The relationship among sequences sampled from this population can be described with a genealogical tree.
This traces the ancestry for a set of lineages backwards in time, and when a pair of lineages share a common ancestor they are said to \emph{coalesce}, which is represented through the two lineages merging in the tree.
The coalescent characterizes the distribution over the shapes of such trees, and the waiting times between coalescence events. When considering a sample of size two, there is only one possible tree topology, so the coalescent fully characterizes the sample genealogy through the distribution of the time of their most recent common ancestor.

\subsection{The coalescent with recombination (CwR)}
While Kingman's coalescent can be effectively applied independently to unlinked loci, nearby loci have highly correlated genealogies, so the effects of recombination must be modeled explicitly.
The CwR was first formulated by \citet{hudson_properties_1983} as a stochastic process backwards in time, which allows for both coalescence and recombination events to occur, resulting in a collection of local trees.
Further developments introduced an alternative graph structure to represent a realization of this stochastic process, where lineages can both merge due to coalescence and split due to recombination
\citep{griffiths_ancestral_1996, Griffiths_two_loucs_arg_1991, Griffiths1997_an_arg}.
This graph structure, known as the ARG, encodes the complete evolutionary history of a sample, including the marginal genealogy at each genomic position \citep{griffiths_ancestral_1996, Griffiths_two_loucs_arg_1991, Griffiths1997_an_arg}.
Rather than representing the CwR as a stochastic process backwards in time, \citet{wiuf_recombination_1999} reframed it as a spatial process that sequentially generates correlated local genealogies along the chromosome.
Under this spatial formulation, the local genealogy at any given position depends on all preceding genealogies, making the process non-Markovian.
This introduces long-range dependencies between genomic regions where non-ancestral segments are ``trapped'' by flanking ancestral segments, rendering the spatial formulation challenging for tractable inference.

The ARG has been described as ``\textit{the holy grail of statistical population genetics}'' because it can fully encode the entire genealogical history of the sample, including the complete record of which lineages experienced recombination and the time when these events occurred \citep{Hubisz2020_holy_grail}.
Given a set of sampled genomes, it would be desirable to infer an ARG that could recover the topologies of local genealogies, the timing of coalescence events, the ages of mutations, and the locations and times of recombination events.
However, the complexity and size of the space of possible ARGs make population genetic inference under the CwR extraordinarily challenging \citep{griffiths_ancestral_1996, mcvean_approximating_2005}.
Even when considering a single locus, the number of possible coalescence topologies for \(n\) sequences is \(\prod_{i=2}^{n}\binom{i}{2}=n!(n-1)!/ 2^{n-1}\) \citep[][Section 3.2]{hein2004gene}, which grows super exponentially, making likelihood calculations for an ARG intractable.
Thus, simplifying the structure of the ARG and approximating the full CwR is necessary for tractable inference.

\subsection{SMC as an approximation to the CwR}
\citet{mcvean_approximating_2005} introduced the SMC as a model to sequentially generate genealogies along a chromosome.
In this sequential view, given a recombination event, the SMC permits the ``floating'' lineage---\ie, the branch below the recombination event---to re-coalesce with any older lineage on the tree except for itself.
Like the CwR, the SMC can also be viewed from a backwards in time perspective, where lineages can only coalesce if they share overlapping ancestral material.
The SMC is thus a Markov process both backwards in time and along the genome, since the genealogy at any given genomic position depends only on the genealogy at the previous position.
Despite this simplification, the SMC generates genealogies with very similar structure, correlation, and patterns of genetic diversity as the CwR, while being much more computationally tractable for inference \citep{mcvean_approximating_2005}.

\begin{figure}[H]
    \centering
    \includegraphics[width=\textwidth]{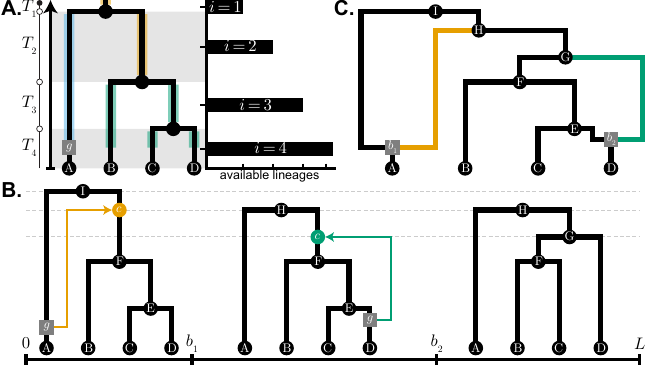}
    \caption{
    \textbf{Illustration of the SMC.} \newline
    \textbf{(A)} Possible types of genealogical transitions due to a recombination event (grey square node \(g\)).
    The “floating” lineage coalesces with one of its ancestral lineages to form the next genealogy along the sequence (see \textbf{B} and \textbf{C}). 
    Type 1 events (sky blue) do not alter the next tree; type 2 events (orange) change the height of the next tree, but not its topology; type 3 events (green) change the topology of the next tree.
    Where the floating lineage coalesces depends on the length of each epoch (\(T_i\), delineated by alternating backgrounds) and the number of ancestral lineages \(i\). 
    The histogram shows the number of lineages \(i\) available for coalescence in each epoch \(T_i\), starting from the time of the recombination event \(g\); the coalescence rate in epoch \(T_i\) is \(\sfrac{i}{2N_e}\).
    \textbf{(B)} SMC process as a sequence of genealogical trees over a genomic region of length \(L\). 
    Each local tree persists until a recombination event (grey square node \(g\)) occurs at a breakpoint \(b_i\).
    The branch between the recombination node and the floating lineage's parent node is then pruned.  
    The genealogy transitions to the next tree along the sequence when the floating lineage is regrafted (colored arrow) via a coalescence event (circle \(c\) node) colored by the types of recombination events described in \textbf{A}.
    Dashed horizontal lines at the height of ancestral nodes I, H and G, highlight that the new genealogies are formed due to the coalescence at the preceding locus. 
    \textbf{(C)} Illustration of an ARG for the corresponding genealogical trees depicted in \textbf{B}.
    Black circle nodes A-D represent sample nodes, and the remaining ancestral black nodes are alphabetically labeled in ascending order based on relative age.
    Grey square nodes denote recombination events and are labeled by their corresponding breakpoint \(b_i\) in \textbf{B}.
    The left and right edges emerging from recombination nodes correspond to the local genealogies to the left and right of breakpoint \(b_i\).
    The orange edge corresponds to the type 2 recombination event resulting in the genealogical transition \([0, b_1) \rightarrow [b_1, b_2)\) and the green edge corresponds to the type 3 recombination event resulting in the genealogical transition \([b_1, b_2) \rightarrow [b_2, L]\).
    }
    \label{fig:smc}
\end{figure}

\citet{marjoram_fast_2006} introduced a modification to the SMC, known as the SMC$^\prime$ (Figure \ref{fig:smc}), which incorporates an additional class of coalescence events, making it a closer approximation to the CwR while still being Markovian.
Specifically, in the sequential view, the SMC$^\prime$ also allows the ``floating'' lineage to re-coalesce with itself, which is not permitted in the original SMC.
From the backwards in time perspective, the SMC$^\prime$ allows coalescence between lineages with overlapping \textit{or} adjacent ancestral material, while the SMC only permits coalescence between lineages with overlapping ancestral material.
This difference means that after a recombination event, the SMC always generates a new marginal genealogy distinct from the previous one, whereas the SMC$^\prime$ allows for self-coalescence that may preserve the previous genealogical structure in the new marginal genealogy \citep{marjoram_fast_2006}.

The key insight of both the SMC and the SMC$^\prime$ models is that by making each local genealogy dependent only on the genealogy at the previous locus, the process becomes Markovian.
This Markovian approximation substantially reduces the space of possible ARGs, providing a much more tractable framework for statistical inference.
Moreover, by studying the joint distribution of coalescence times in a two-locus Markov chain model, \citet{Wilton2015_smc_accuracy} demonstrated that both the SMC and SMC$^\prime$ are good approximations to the full ARG, with the SMC$^\prime$ being notably more accurate.
Furthermore, they showed that the joint distribution of pairwise coalescence times at sites flanking recombination breakpoints under the SMC$^\prime$ is identical to that under the full ARG \citep{Wilton2015_smc_accuracy}.
For these reasons, the SMC$^\prime$ has become the more popular choice for inference methods, and throughout this review, when we refer to ``the SMC'' and ``SMC-based methods'' we are, more formally, discussing the SMC$^\prime$.
We describe the SMC$^\prime$ algorithm in Algorithm \ref{alg:smcprime}; we note that reversing the order of steps 4 and 5 yields the original SMC formulation described in \citet{mcvean_approximating_2005}.

\begin{algorithm}[!htbp]
\caption{
    Generating genealogies under the SMC$^\prime$ \citep{marjoram_fast_2006}.
} \label{alg:smcprime}
\DontPrintSemicolon
\textbf{Input:} Effective population size \(N_e\), sequence length $L$, and per-site per-generation recombination rate \(r\). Compute the population-scaled recombination rate \(\rho = 4N_e r\).\;

\textbf{1.} Set the first left interval position as \( x = 0 \) and generate a coalescent tree for \( x \) under Kingman's coalescent, denoted as \( \mathcal{T}(x) \). Denote the length of the tree at \( x \) as \( \mathcal{L}(x) \), which sums all of the branch lengths in the tree.\;
\textbf{2.} Generate the right interval position as \( y \sim \text{Exp}(\frac{\rho}{2} \mathcal{L}(x)) \), the distance along the chromosome until the next recombination event---\ie, tree \( \mathcal{T}(x) \) spans the genomic interval \([x, x+y)\). \;
\textbf{3.} Pick a point \( g \) on the tree \( \mathcal{T}(x) \) uniformly.\;
\textbf{4.} Add a recombination event to the tree \( \mathcal{T}(x) \) at point \( g \), resulting in a graph, which occurs at chromosomal location \( x + y \). The left emerging branch follows the original path in \( \mathcal{T}(x) \), while the right emerging branch re-coalesces at some point \(c\) higher up on the graph (after a waiting time which is exponentially distributed, with rate proportional to the number of ancestral lineages at each epoch).\;
\textbf{5.} Delete the part of the left emerging branch that lies between the newly added recombination event at point \( g \) and the next coalescence event along the branch, thus reverting the graph back to a tree.  \;
\textbf{6.} Set the new left interval position as \( \hat{x} = x + y \), where \( \mathcal{T}(\hat{x}) \) and \( \mathcal{L}(\hat{x}) \) denote the new tree constructed starting from position \( \hat{x} \) and its associated tree length.\;
\textbf{7.} \If{\( \hat{x}< L \)}{
Set \(x = \hat{x}\) and return to step 2.\;
}

\end{algorithm}

\subsection{Statistical inference using the SMC}
\label{sec:primer:hmm}
The most common application of the SMC is to estimate changes in \Ne{} over time from sequencing data.
To infer model parameters under the SMC conditional on the data, methods such as \texttt{PSMC} or \texttt{MSMC} \citep[and many others reviewed in Sections \ref{sec:vector} and \ref{sec:partitioning};][]{li_inference_2011, schiffels_inferring_2014} use a Hidden Markov Model (HMM).
HMMs are a powerful and flexible inferential framework; indeed, one of the most useful features of the SMC approximation to the CwR is that it is Markovian and therefore allows the use of HMMs---for more about HMMs in this context, see \citet{durbin_biological_1998}.

\begin{figure}[H]
    \centering
    \includegraphics[width=\textwidth]{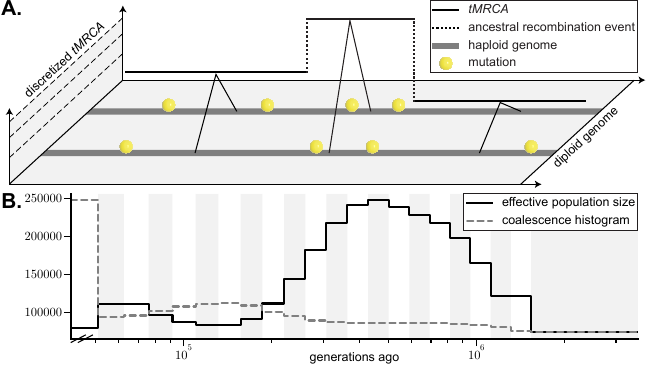}
    \caption{
    \textbf{Overview of \texttt{PSMC} framework.} \newline
    \textbf{(A)} \texttt{PSMC} is a Hidden Markov Model (HMM) used to infer piecewise constant historical population sizes from the spatial distribution of heterozygous sites between two haploid sequences (grey lines on the x-axis plane).
    Hidden states are discrete intervals of times to the most recent common ancestor (\textit{tMRCA}), shown as dashed lines on the y-axis plane.
    Transitions between hidden states represent changes in the \textit{tMRCA} of the two sequences, reflective of ancestral recombination events (dotted vertical lines).
    Observations consist of the sequence of diploid genotypes at each site, where a yellow sphere indicates a heterozygous site (\ie{}, the mutation occurred after the \textit{tMRCA}) and its absence indicates a homozygous site.
    Emissions correspond to the probability of observing these mutations given a \textit{tMRCA} for the two sequences.
    \textbf{(B)} Using the Baum-Welch algorithm to estimate model parameters, \texttt{PSMC} reconstructs a piecewise constant ancestral population size history.
    The crux of \texttt{PSMC}'s inference relies on the inverse relationship between the frequency of coalescence events (grey dashed line) in each time interval (alternating background) and the effective population size (black line) during that time period.}
    \label{fig:psmc}
\end{figure}

\texttt{PSMC} models the density of heterozygous sites along the genome to infer the coalescence time at every genomic position using sequence data from a single diploid individual (Figure \ref{fig:psmc}).
Since only two sequences are analyzed at a time, each genealogical tree along the sequence contains only two lineages and one coalescence time---\ie, the time to the most recent common ancestor (\textit{tMRCA}) for a diploid individual's two homologous chromosomes.
Time---the axis along which lineages in each tree coalesce backwards in time from the present---is partitioned into discrete time intervals to form the \emph{hidden state space} of a discrete HMM.
The \emph{observations} in the data are whether or not alleles match (\ie, the genotype is homozygous or heterozygous) at each site in the sequence, where a heterozygous site indicates that a mutation arose more recently than the \textit{tMRCA} of the individual's two haploid lineages.
The \emph{emission probabilities}, based on the infinite sites mutation model, describe the probability that the observation is homozygous or heterozygous, given that the lineages coalesced in a particular time interval, and are determined by the population-scaled mutation rate \(\theta = 4N_e \mu\).
The \emph{transition probabilities}, derived from the SMC framework, describe the probability that a locus whose lineages coalesced in one particular time interval is adjacent to a locus whose lineages coalesced in the same or different time interval, and are governed by the population-scaled recombination rate \(\rho = 4N_e r\) and the given evolutionary model.

\texttt{PSMC} uses the Baum-Welch algorithm---a special case of the expectation-maximization algorithm---to estimate its parameters, namely the piecewise constant coalescence rates and the population-scaled recombination rate \(\rho\).
In the expectation step, the forward-backward algorithm computes the posterior probability of coalescence in each hidden state for each observation, this posterior distribution is then used to construct the expected transition matrix.
In the maximization step, the expected transition matrix is used to update the model parameters.
The updated parameters are subsequently used to recompute the expected transition matrix, and the process iterates until convergence.
After convergence is achieved, \Ne{} is estimated from the \texttt{PSMC}'s piecewise constant coalescence rates as the inverse of the coalescence rate, after appropriately scaling the time units using the per-site per-generation mutation rate \(\mu\) and generation time.

\subsubsection{Inferring demographic histories}
The ability to infer changes in historical population sizes from a single diploid genome sequence has made \texttt{PSMC} a very commonly used tool in the field of evolutionary genetics \citep{leon_hilgers_avoidable_2025}. 
Building on this foundation, several methods have extended \texttt{PSMC}'s capabilities to improve inference of historical population sizes over time. The multiple sequentially Markovian coalescent (\texttt{MSMC}) was introduced as an extension that takes multiple (typically as many as eight) diploid genome sequences as input \citep{schiffels_inferring_2014}.
\texttt{MSMC} estimates the time of the first coalescence event among all input sequences, offering two key advantages over its predecessor: for sequences sampled from the same population, the larger sample size enables greater resolution in population size inference for more recent time periods; for sequences sampled from two different populations, it infers the relative cross coalescence rate---a measure of genetic separation between populations based on the ratio between cross-population and within-population coalescence rates.
\texttt{MSMC} also employs a more accurate approximation to the CwR \citep[\ie, SMC$^\prime$]{marjoram_fast_2006} compared to the original \texttt{PSMC} implementation \citep{li_inference_2011, mcvean_approximating_2005}.
Building on \texttt{MSMC}, \texttt{MSMC2} was introduced as a simpler yet more powerful approach \citep{wang_tracking_2020}.
Rather than inferring the first coalescence event among input sequences, \texttt{MSMC2} runs \texttt{PSMC} on all possible pairs of sequences and uses composite-maximum likelihood to infer model parameters.

Another class of methods leverages the conditional sampling distribution (CSD), which describes the probability of observing a new haplotype given a set of previously observed haplotypes and a demographic model \citep{Paul2011_csd_smc, Paul2010_csd, Steinrucken2013_csd_smc_structured}.
\texttt{diCal} employs the CSD to infer recent population size history using up to 10 genomes \citep{Sheehan2013_dical}, while \texttt{diCal2} extends this framework to infer more complex demographic histories involving population splits, admixture, and migration \citep{Steinrucken2019_dical2}.
\texttt{SMC}$_{\text{\texttt{++}}}$ builds on these approaches by combining the CSD with explicit modeling of the site frequency spectrum (SFS), enabling analysis of hundreds of unphased genomes \citep{terhorst2017robust}.

Several other methods have introduced novel approaches to infer population size histories under the SMC framework.
\texttt{SMCSMC} uses sequential Monte Carlo to estimate the posterior distribution of the hidden continuous time Markov process from four diploid human genomes, and avoids biases in the expectation-maximization algorithm by employing variational Bayes to model uncertainty in rare events \citep{Henderson2021_smcsmc}.
\texttt{CHIMP} scales to hundreds of genomes by modeling the latent space as either the height or length of the tree, obtaining the Markov transition probabilities through numerically solving systems of differential equations, and using composite likelihood to scale to large sample sizes \citep{Upadhya2022_chimp}.
Additionally, a Gaussian process-based Bayesian nonparametric method has been developed that avoids discretization of the parameter space while providing uncertainty estimates for inferred parameters \citep{Palacios2015_gp_ne}.

\subsubsection{Inferring coalescence times to study natural selection}
While SMC-based methods have primarily focused on inferring demographic histories, the sequence of coalescence times inferred along the genome also provides valuable information for studying natural selection. 
For example, regions under recent positive selection are enriched for young \textit{tMRCA}s, whereas regions under long-term balancing selection are enriched for ancient  \textit{tMRCA}s.

\texttt{PSMC}'s posterior decoding scales quadratically in the number of hidden states, making it impractical to apply pairwise to a large number of sequences. By exploiting symmetries in the discretized SMC transition process, \citet{Harris2014_linear} showed that the \texttt{PSMC} algorithm can be decoded in time linear. The Ascertained Sequentially Markovian Coalescent \citep[\texttt{ASMC};][]{palamara2018high} utilizes this linear decoding, combined with efficient dynamic programming, to run pairwise \textit{tMRCA} inference on biobank scale data, finding novel signatures of recent positive selection and illuminating the landscape of selection on complex traits.
Further improving the efficiency of \texttt{PSMC}'s posterior decoding  \citet{schweiger_ultrafast_2023} introduced \texttt{Gamma-SMC}.
Whereas \texttt{ASMC}'s decodes in linear time, \texttt{Gamma-SMC} decodes in constant time. This speedup is accomplished by representing the hidden distribution over the coalescence times with a Gamma distribution, and computing a forward/backward pass through the observed data by using a flow-field to approximate the transition density. Furthermore, the gamma representation allows the \textit{tMRCA}s to be modeled in continuous time and is thus more robust to misspecification.

\section{SMC with Joint Parameter Inference} 
\label{sec:vector}
\begin{figure}[H]
    \centering
    \includegraphics[width=\textwidth]{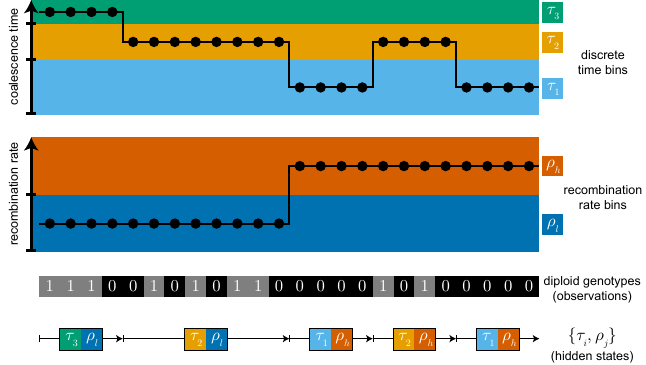}
    \caption{
    \textbf{Illustrative example of joint parameter inference.} \newline
    The Hidden Markov Model (HMM) infers both a discretized coalescence time history (vector of \(\tau_i\)) and a discretized recombination rate map (vector of \(\rho_h\) and \(\rho_l\), indicating high and low recombination respectively). 
    Hidden states are tuples of all combinations of coalescence time bins \(\tau_i\) and recombination rate bins \(\rho_j\). 
    An HMM path is depicted underneath the observed diploid genotypes (“0” for homozygous and “1” for heterozygous), which corresponds to a sequence of hidden states (pairs of \(\{\tau_i, \rho_j\)\}) emitted by the sequence of observations.}
    \label{fig:vector}
\end{figure}

Several extensions of the SMC framework can infer not just a vector of inverse coalescence rates (\ie, \Ne) through time, but also vectors of other biological parameters of interest, including mutation and recombination rates, mating system properties, and sequence constraint (Figure \ref{fig:vector}).

\subsection{SMC-based inference with mutation and recombination rate heterogeneity}
\label{sec:vector:iSMC}
\texttt{PSMC} brought large inferential power to genome projects by allowing the estimation of historical population sizes from a single unphased diploid genome. 
The integrative Sequentially Markovian Coalescent (\texttt{iSMC}) extends what can be learned from such data by extending the SMC framework to also estimate a recombination map along the genome  \citep{barroso_inference_2019}. 
In the original \texttt{PSMC} framework, the transition rates between coalescent time bins assumes a constant, global recombination rate parameter. 
\texttt{iSMC} models spatial variation in recombination rate using a Gamma distribution with \(k\) discretized categories, allowing for the inference of local recombination rates along the genome.

\texttt{iSMC} is not only able to distinguish between the signals of variable population size and recombination rate in SNP data, but provides more accurate estimates when inferring both concurrently. 
Although \texttt{iSMC} produces lower resolution recombination maps than linkage disequilibrium (LD) based methods, \texttt{iSMC} is more accurate when only a small number of samples are available, or in the presence of historical population size variation \citep{barroso_inference_2019,Dutheil2024}. 
However, \texttt{iSMC} introduces biases if the number of recombination categories $k$ is poorly selected \citep{Dutheil2024}. 
\texttt{iSMC} is also limited to estimating the background recombination rate and cannot detect finer-scale variation such as recombination hotspots. 
Further development is needed for small genomes and those with high gene conversion rates \citep{Dutheil2024, Schweizer2021}.

\texttt{iSMC}'s ability to analyze single genomes has enabled the estimation of recombination maps for individuals from diverse human populations, as well as for archaic hominins like Neanderthals and Denisovans \citep{barroso_inference_2019}. These findings show that the recombination landscape divergence aligns closely with the divergence of both extant and extinct lineages.
This framework has also enabled  the first recombination map estimates for a wide range of non-model and threatened species \citep{Robinson2021, Cui2024, daFonseca2024, nouhaud_rapid_2022, Unneberg2024}.
These recombination maps have helped highlight the significance of genome-wide recombination rate variation in local genome evolution.  
For example, \texttt{iSMC} has been used to show that for conservation efforts, low-recombining genomic regions are of particular concern, as they are more vulnerable to the effects of linked selection.
Directional selection reduces \Ne{} at linked sites due to genetic hitchhiking \cite[\ie{} Hill-Robertson effect;][]{Charlesworth1993_effect_of_del_muts, Nordborg1996_recomb_and_bgs}, reducing local levels of genetic diversity proportionally to the strength of selection and inversely to the local recombination rate \citep{Charlesworth2009Ne}.
For regions of the genome with lower recombination rates, this effect is particularly pronounced, where linked selection can create long-lasting reductions in both \Ne{} and genetic diversity.
Low-recombining regions were more likely to contain runs of homozygosity, a common conservation measure used to indicate inbreeding, in the critically endangered Chinese \textit{Bahaba} fish \citep{Cui2024}. 
These regions were also more likely to contain high-impact substitutions and highly disruptive variants. Higher recombination rates in smaller chromosomes and distal regions of larger chromosomes helped explain the increased heterozygosity in these regions in the critically endangered California condor \citep{Robinson2021}. 
Low-recombining regions in small pelagic European sardines also contained increased genetic differentiation between populations \citep{daFonseca2024}. 
However, purging of deleterious alleles, a common advantage to prolonged small \Ne{}, appears to be more efficient in low-recombining regions in hybrid ant genomes \citep{nouhaud_rapid_2022}. 

\texttt{iSMC} has been extended to estimate mutation rates across the genome alongside demographic histories and recombination rates \citep{Barroso2023}. 
This approach helps partition the relative contributions of genetic drift, linked selection, and local mutation rates to the evolution of genetic diversity. 
For instance, this extension revealed that local mutation rates primarily drive patterns of genetic diversity in \textit{Drosophila melanogaster}, with linked selection playing a secondary role \citep{Barroso2023}.
While \texttt{iSMC} models heterogeneity in mutation rate across the genome as additional hidden states using a Gamma distribution with \(m\) discretized categories in an HMM, other strategies have been developed to overcome \texttt{PSMC}'s assumption of a genome-wide constant mutation rate.

\texttt{PSMC+} was introduced by \citet{cousins_accurate_2024} to model background selection. 
A classic model of the effect of background selection on pairwise diversity re-scales the effective population size by a factor \(B\) to account for loss of diversity due to linkage with alleles experiencing purifying selection \citep{charlesworth1993effect}. 
As mentioned in Section \ref{sec:primer:hmm}, \texttt{PSMC}'s emission probabilities are governed by the population-scaled mutation rate, which is assumed to be constant along the genome, and describes the probability of observing a mutation given a coalescence time.
\texttt{PSMC+} modifies \texttt{PSMC}'s emission model to account for local variation in mutation rate by scaling \(\theta\) by a factor of \(f_i\) for each observation \(i\), which reduces bias for inferring \Ne{} if the true scaling factor is known.
In doing so, \texttt{PSMC+}'s emission probabilities now describe the probability of observing a mutation given a coalescence time \textit{and} local mutation rate scaling factor.
The authors show via simulations that this procedure improves the accuracy of estimates of \Ne{} through time in the presence of background selection. 
Using a simple \(B\)-map based on distance to exons, or simply rescaling \textit{post-hoc}, performed similarly to the high-resolution Murphy $B$-map in humans, suggesting that minimal prior knowledge of background selection is required. 
When not accounting for background selection, the ratio of effective population sizes between X-chromosomal and autosomal regions varies substantially over time, although this could be due to life history or mutation rate variation \citep{cousins_accurate_2024}.
It should be noted that while \citet{cousins_accurate_2024} primarily focused on the effect of background selection, \texttt{PSMC+}'s emission model can be used to account for any local variation in mutation rate.

\subsection{Detecting life history traits with the SMC}
SMC-based methods inherit neutral assumptions from Kingman's coalescent (Section \ref{sec:primer:coal}). 
However, many systems---especially plants and invertebrates---exhibit ecological life history traits that violate two assumptions of this model: sexual reproduction through random mating, and non-overlapping generations. 
Recent extensions to the SMC framework have addressed these challenges, yielding effective models for phenomena such as self-fertilization and seed- (or egg-) banking. 

Self-fertilization is a non-random mating system prevalent in plant and fungal species in which an individual's gametes are more likely to fuse with its own than those of another individual \citep{Nordborg2000-selfing-arg}. 
Self-fertilization reduces genetic diversity and thus decreases the effective population size,
resulting in genealogical trees with shorter branches and fewer mutations and recombination events than those generated under the standard SMC \citep{Nordborg2000-selfing-arg}. 
Seed (or egg) banks are an adaptation to unpredictable environments whereby seeds remain dormant for multiple generations before germinating \citep{Cohen1966-seedbanks-bh}.
Seed banks break the assumption of non-overlapping generations by allowing past generations to contribute genetically to the present. 
This increases the effective population size,
resulting in genealogical trees with longer branches and more mutations and recombination events than those generated under the standard SMC \citep{Lennon2021-seedbak-theory-review}.

To explicitly infer these life history traits alongside \Ne, \citet{sellinger_inference_2020} developed the ecological sequentially Markovian coalescent (\texttt{eSMC}). 
The \texttt{eSMC} can be viewed as a rescaled extension of \texttt{PSMC}: it leverages the fact that self-fertilization and seed banking alter the recombination and mutation processes in opposite directions and distort the ratio of recombination to mutation rates. 
By exploiting this deviation, the \texttt{eSMC} is able to jointly infer a piecewise constant demographic history along with either a fixed self-fertilization rate or a germination rate. 

Applying \texttt{eSMC} to Swedish and German \textit{Arabidopsis thaliana} populations, \citet{sellinger_inference_2020} showed that self-fertilization rates introduce minimal bias into historical \Ne{} estimates compared to traditional SMC-based inference methods \citep{Durvasula2017_arabidopsis}.
Moreover, the \texttt{eSMC} rejected the previously-proposed existence of seed banks in these populations. 
In contrast, when applied to \textit{Daphnia pulex}, \texttt{eSMC} inferred an egg bank lasting 3–18 generations, consistent with empirical observations \citep{Brendonck2003_egg_bank_gen_time}.
A standard SMC approach neglecting egg-banking would have reported a biased \Ne{} trajectory, and obscured the role of egg banks in maintaining genetic variation.

The \texttt{eSMC} has also been extended to allow for the joint inference of historical population sizes, recombination rates, and selfing (or germination) rates by allowing these parameters to vary through time \citep[\texttt{teSMC};][]{strutt_joint_2023}. 
This enables the dating of transitions from outcrossing to selfing based on temporal changes in the ratio of recombination to mutation rate.
Returning to \textit{A. thaliana}, \citeauthor{strutt_joint_2023} estimated that a transition to self-fertilization occurred approximately 600kya. 
Since \texttt{teSMC} can distinguish historical changes in \Ne{} from changes in self-fertilization rates, it also detected a previously-unrecognized population decline coinciding with the transition from outcrossing to selfing.
This demonstrates how incorporating time-varying vectors of ecologically-relevant parameters can enhance the resolution of SMC-based inference and our understanding of mating system evolution.

\texttt{eSMC} has also been extended to accommodate systems where a few individuals produce a large proportion of offspring \citep[\texttt{SM$\beta$C};][]{Korfmann2024}. 
This can be caused by skewed offspring distributions---as are seen in plants, invertebrates, prokaryotes, and fish---or strong selection events. 
\texttt{SM$\beta$C} employs the \(\beta\)-coalescent to allow long range dependencies between coalescent trees along the genome, which are indicative of multiple merger events. 
\texttt{SM$\beta$C} has been shown to distinguish between the influences of skewed offspring distribution and positive selection while simultaneously modeling historical population sizes to a high degree of accuracy \citep{Korfmann2024}. 

\section{SMC with Lineage Partitioning}
\label{sec:partitioning}
\begin{figure}[H]
    \centering
    \includegraphics[width=\textwidth]{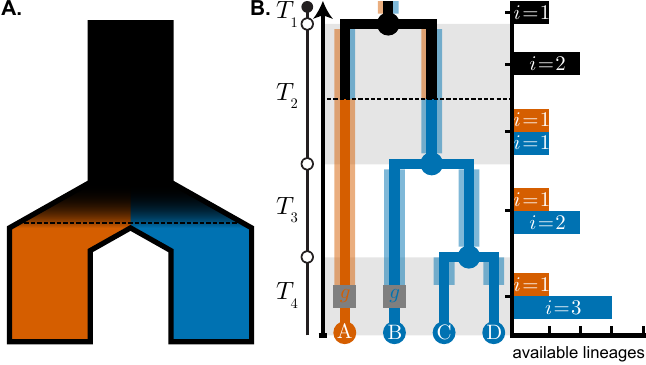}
    \caption{
    \textbf{Illustrative example of lineage partitioning.} \newline
    \textbf{(A)} Lineages are partitioned into colored classes (vermilion or blue) where cross-color coalescence is restricted until after the barrier (black dashed line) arose.
    \textbf{(B)} Possible genealogical transitions following a recombination event (grey square node \(g\)) stratified by colored class (vermilion or blue \(g\)).
    Starting from the time of the recombination event, the shading behind each branch indicates the colored class identity of the ancestral lineage available for coalescence.
    With respect to each colored class, where the floating lineage coalesces depends on the duration of each epoch (\(T_i\), delineated by alternating backgrounds), and the number of ancestral lineages from the same colored class available during that epoch.
    The histogram shows the number of ancestral lineages \(i\) available for coalescence in each epoch stratified by colored class, \(T_i\), which is proportional to the coalescence rate \(\sfrac{i}{2N_e}\). 
    }
    \label{fig:partitioning}
\end{figure}

Under the original SMC model, any lineage may undergo recombination, and that lineage is permitted to re-coalesce with itself or any (older) lineages with overlapping or adjacent ancestral material in the current genomic interval.
However, it is possible to define an SMC-like process where lineages are partitioned into different labeled classes---\eg, by carrier status for a chromosomal inversion, or membership of separated populations. 
In these models, lineages are imagined to have different colors throughout time, where the color is reflective of the lineage's class membership at any given time.
Recombination and/or coalescence events can then be modulated to occur at different rates between lineages of the same or different color.
We refer to this type of modification as the SMC with lineage partitioning (Figure \ref{fig:partitioning}).

\subsection{Modeling and detecting structural variation with the SMC}
\label{sec:partitioning:svs}
\citet{peischl_sequential_2013} introduced an SMC variant that uses a lineage partitioning approach to model polymorphic inversions. 
Each lineage is assigned to a specific color---standard or inverted---and coalescence events are restricted to occur only within a color. 
While coalescence is not allowed between lineages of different colors, to model both homokaryotypic recombination and gene flux (\ie{}, heterokaryotypic recombination) events, recombination events are not strictly limited to only lineages of the same color, albeit at a drastically reduced rate than same-color recombination events.
Separating lineages by color reflects the underlying biology of inversions, where recombination is highly suppressed between inversion carriers and non-carriers.

Although this model has only been derived for a single population, and not yet applied to any empirical data, it has generated novel results about LD.
\citet{peischl_sequential_2013} studied the correlation in coalescence times at pairs of sites across an inverted region generated by their method.
They found that, unlike standard models of non-inverted regions, where LD decays with the physical distance between sites, an inverted region can maintain long-distance associations between sites. 
This pattern of LD arises because gene flux events are effectively migration events between the standard and inverted classes of lineages, breaking down the associations between sites. 
However, unlike migration events, the rate of gene flux events varies spatially along the inversion.
Gene flux is more likely to occur in the interior of the inversion, and less likely near the breakpoints. 
By extending the SMC to model inversions, \citet{peischl_sequential_2013} demonstrated that the strength of LD between sites on an inversion depends on both the physical distance between sites and their spatial location on the inversion.

\citet{ignatieva_length_2024} used the SMC to identify and analyze structural 
variants (SVs) in empirical data, by looking for distortions in the genealogy due to the localized suppression of recombination that is induced by SVs.
The crux of the method (implemented in the software tool \texttt{DoLoReS}) is that, in the case of an inversion, recombination is suppressed in individuals with only one copy of the inverted orientation. 
If a recombination event occurs on a lineage within the clade of chromosomes carrying the inversion, it is highly likely to coalesce with another lineage within the clade of inverted samples. 
Similarly, if a recombination event occurs on a branch that does \emph{not} harbor the inversion, it will likely \emph{not} coalesce with any lineages in the clade of inverted samples. 
Consequently, samples that harbor an inversion should persist as a clade in the ARG for longer stretches of the genome than would otherwise be expected. 

To identify such signals of locally suppressed recombination, \citet{ignatieva_length_2024} used the SMC to derive the distribution of the genomic span for a given clade, conditional on the local tree where the clade first appears. 
For a given input genealogy, \texttt{DoLoReS} then computes the genomic span of each observed clade, and uses the derived distribution to compute a \(p\)-value which indicates whether the clade has a longer genomic span than expected.

Using a genealogy reconstructed from human sequencing data \citep{speidel2019method}, \texttt{DoLoReS} identified 50 regions of the genome with strong signals of suppressed recombination. 
This includes the well-known inversion on 17q21.31, and a known region of complex structural variation on 16p12.2, as well as a novel 760kb inversion on 10p22.3 which is common in South Asian populations---and which reads-based methods have failed to detect due to lack of strong signal from reads spanning the inversion breakpoints. The method also identified a number of regions with locally suppressed recombination that did not show strong evidence of structural variation; further analysis revealed that the observed signals may arise through expression-dependent suppression of crossovers within genes expressed in meiosis.

\subsection{SMC-based inference with population structure}
A widely-adopted method, \texttt{MSMC-IM} \citep{wang_tracking_2020} uses a lineage partitioning strategy to model lineages from two populations and infer a time-dependent migration rate estimate between them.  
Input sequences are colored by which of the two populations they were sampled from. 
Rather than restricting coalescence between lineages of different colors, as in the case of inversions \citep{peischl_sequential_2013}, \texttt{MSMC-IM} expects fewer coalescence events to happen between lineages from different populations, as these events reflect migration.
\texttt{MSMC-IM} uses the SMC-based method \texttt{MSMC2} \citep{malaspinas_genomic_2016, wang_tracking_2020} to infer three separate rates of coalescence: within each population and between the two populations.
It then uses these three histories to fit an isolation-migration model \cite[IM; based on][]{hobolth_computing_2011}.
The fitting process effectively converts the three coalescence rates inferred from \texttt{MSMC2} into effective population size histories for each population and the migration rate between them. \texttt{MSMC-IM} has been applied to both intra- and inter-specific questions \citep{lescroart_felids_2023, crossman_whales_2024, Wang_polar_2022}, leading to novel population genomic and phylogeographic inference. 

Mutation rate and generation time are key parameters for SMC interpretation, yet not always empirically estimated for non-model species. 
One such species is the mosquito \textit{Aedes aegypti} for which a human specialist \textit{Ae.\ aegypti aegypti} is a major vector for several viral pathogens. 
\texttt{MSMC-IM} was used to understand the evolution of the human specialist population from the generalist \textit{Ae.\ aegypti formosus}; however, the authors first needed to estimate these key parameters. 
To do this they created \texttt{MSMC-IM} curves from African and South American populations of \textit{Ae.\ aegypti aegypti}, and calibrated the curve based on the timing of the human slave trade \citep{rose_dating_2023}. 
Their calibration enabled them to identify that deep population structure in Africa accelerated as the landscape dried and water became associated with human settlements, thereby giving rise to the human specialist. 
Additional sampling from this system identified the phylogeography of global range expansion of this anthrophilic vector \citep{crawford_dengue_2024}.

\texttt{MSMC-IM} is being applied to understand hybridization and introgression. 
While unidirectional introgression from polar into brown bears has been previously inferred, the timing was unknown. 
Analysis of the migration rate curves from \texttt{MSMC-IM} detected the timing of speciation (500kya - 400kya), and a previously unidentified ancient hybridization affecting global brown bear populations \citep[100kya;][]{Wang_polar_2022}. 
Similarly, \texttt{MSMC-IM} signatures between wolf and coyote populations identified both speciation and recent hybridization with on-going gene flow that aids in defining evolutionary units for conservation \citep{vilaca_canids_2023}. 

The use of \texttt{MSMC-IM} has been limited for questions of selection, but has supported downstream analyses. 
For example, \cite{puckett_genetic_2023} inferred pairwise migration rates among four populations of American black bears to parameterize a set of \texttt{SLiM} \citep{haller_slim3_2019} models with varying selection coefficients. 
They found contemporary allele frequencies of the causative locus for brown coat color were best described under a model with a low but positive selection coefficient. 
\citet{wei_tomato_2023} compared the population divergence time due to colonization of the Andes mountains of an endemic tomato species, \textit{Solanum chilense}, to the age of selective sweeps within the population. 
They found that sweeps in the highlands population were generally closer to the time of colonization of the novel environment than in the age of sweeps in lowland populations.

\citet{Shchur2022_misti} introduced \texttt{MiSTI} as an alternative SMC-based approach to account for population structure, that can infer changes in historical effective population sizes for two populations while correcting for the confounding effects of migration. 
The method combines coalescence rates inferred from \texttt{PSMC} with the joint site frequency spectrum (jSFS) from two diploid individuals to fit a time-dependent migration model.
By incorporating both the \texttt{PSMC} results and the jSFS, \texttt{MiSTI} can disentangle the complex interplay between population size changes and migration, providing estimates of migration-corrected historical population sizes, split time between the populations, and asymmetric migration rates.
The crux of \texttt{MiSTI}'s approach lies in its modeling distinctions between different conceptualizations of effective population sizes. 
Specifically, \citet{Shchur2022_misti} modeled the distinction between the \textit{ordinary} effective population size---\ie, as inferred by methods like \texttt{PSMC}---and the \textit{local} effective population size. 
The \textit{ordinary} effective population size of an admixed population is comprised of the \textit{local} effective population sizes of the parental populations, which explains why \texttt{PSMC} results can be biased in the presence of population structure \citep{Mazet2016_iicr}.
In contrast, the \textit{local} effective population size represents the effective population size considering only unadmixed individuals within a parental population---\ie, the inverse coalescence rate conditional on both lineages belonging to the same population at a given time.
By modeling the relationship between these two types of effective population sizes as a function of the split time  and post-separation migration between the two populations, \texttt{MiSTI} can disentangle the effect of migration on changes in historical population sizes over time.

\citet{Steinrucken2018_dical_admix} developed \texttt{diCal-ADMIX}, an SMC-based method for identifying genomic regions resulting from introgression---the transfer of genetic material between previously isolated populations through hybridization and subsequent back-crossing with one of the parental populations. 
This model-based approach analyzes three populations: a target population (\ie, the proposed recipient of introgression), a putative source population, and a control population presumed to have no genetic contribution from the source population. 
\texttt{diCal-ADMIX} implements an HMM based on the CSD \citep{Paul2011_csd_smc, Paul2010_csd, Steinrucken2013_csd_smc_structured} and requires a pre-specified demographic model describing the joint evolutionary history of all three populations.
The crux of this method relies on the CSD framework: at each locus, haplotypes from the control and source populations form two unchanging ``trunk'' genealogies, while target population haplotypes are iteratively sampled and absorbed by one of these trunk haplotypes. 
When iteratively sampling the target haplotypes, lineages belonging to different populations cannot coalesce unless a gene flow event occurs with a given rate at some point in the past. 
During a gene flow event, the target haplotype is absorbed by one of the two trunk genealogies, where the underlying dynamics of this absorption process are governed by the demographic model.
Thus, the trunk genealogy and time of absorption for a target haplotype can change from locus to locus, reflective of the fact that target haplotypes are realized as a mosaic of the source and control populations.
\texttt{diCal-ADMIX} models this absorption process for an additional target haplotype, with hidden states representing both the potential absorbing trunk populations and the timing of absorption, and transition and emission probabilities previously defined for the structured SMC-CSD with migration \citep{Steinrucken2013_csd_smc_structured}. 
After decoding the HMM, the method marginalizes over absorption times and groups target haplotypes by trunk population, resulting in the probability that each target haplotype derives from either the control population or was introgressed from the source population for each locus.
When applied to recover Neanderthal introgressed regions in European and East Asian populations (using African populations as the control), \texttt{diCal-ADMIX} confirmed the well-established pattern of reduced introgressed ancestry on the X chromosome compared to autosomes. 
Through subsequent gene enrichment analyses and simulations, \citet{Steinrucken2018_dical_admix} demonstrated that this depletion of Neanderthal ancestry is more consistent with the mutational load hypothesis (\ie, higher genetic load in Neanderthals due to smaller effective population size) rather than Dobzhansky–Müller incompatibilities (\ie, invoking reproductive isolation) between archaic and modern humans.

In contrast to the above methods that infer gene flow between given populations, \citet{cousins_structured_2025} introduced \texttt{cobraa}, a method that detects the presence or absence of gene flow from an unsampled population using just a diploid sequence. \texttt{cobraa} is an extension of the \texttt{PSMC} algorithm that explicitly incorporates a model of population structure in the transition matrix of its HMM. This model of population structure assumes the given population was panmictic until time \(T_1\), at which point a fraction \(\gamma\) of lineages instantaneously migrate to an unsampled population.
This unsampled population remains isolated until time \(T_2\), after which all lineages from both populations merge into a single panmictic ancestral population. Using the Baum-Welch algorithm (see Section \ref{sec:primer:hmm}), \texttt{cobraa} infers both the historical population size changes for the sampled population, the admixture fraction \(\gamma\), the divergence \(T_2\) and admixture \(T_1\) times, and the size of the unsampled population.
Importantly, \texttt{cobraa} does not assume \textit{a priori} that the input diploid sequence was sampled from a structured population.
Instead, it infers a structured model and an unstructured model (as in \texttt{PSMC}) and compares the log-likelihoods between them to determine which better explains the data.
When the structured model provides a better fit, \citet{cousins_structured_2025} introduced a complementary HMM, \texttt{cobraa-path}, which is conceptually similar to \texttt{cobraa} but with hidden states corresponding to both the discrete coalescence time intervals and ancestral lineage path, with modified transition and emission probabilities.
The key innovation of \texttt{cobraa-path} is its ability to infer local ancestry states along the genome, identifying regions where none, one, or both lineages traced their ancestry through the unsampled population.
In their analysis, \citet{cousins_structured_2025} found that a model where an ancestral population diverges ~1.5Mya and subsequently admixes ~300kya with an unsampled population in a ratio of 80:20\% fits human genomic data substantially better than a model without deep admixture.
Furthermore, after inferring the genomic regions derived from this admixture event, the authors discovered a negative relationship between this admixed ancestry, the distance to the closest coding sequence, and the high-resolution \textit{B}-map (which quantifies the strength of background selection).
They suggest this pattern results from the purging of the unsampled population's ancestry following the admixture event.

\subsection{Using the SMC to study speciation}
While most SMC-based methods focus on studying the evolutionary history of a single population, several approaches have adapted the SMC framework to study species-level relationships. These methods employ a lineage partitioning strategy defined by the underlying species tree, where lineages are colored by species identity and speciation events create barriers to coalescence. 

This strategy was first implemented in \texttt{CoalHMM} \citep{hobolth_genomic_2007, Dutheil2009_coalhmm_smc}, which is parameterized by a three-taxon species tree.
Under this model, there are four possible coalescent histories: one lineage sorting, where lineages from the two sister species coalesce (or sort) into their most recent common ancestral species, and three incomplete lineage sorting (ILS), which are genealogically discordant with the species tree.
\texttt{CoalHMM} models the sequence of local coalescent histories from a three-species alignment, with ancestral states polarized by an additional outgroup species.
Unlike standard phylogenetic methods that infer divergence times between species using genomic divergence and fossil calibrations \citep{Yang2007_paml, Ronquist2012_mrbayes, Bouckaert2019_beast}, \texttt{CoalHMM} directly infers speciation times---which typically predate divergence times---and ancestral effective population sizes by leveraging information from the local coalescent histories constrained by the species tree.
However, \texttt{CoalHMM} yields biased parameter estimates for the three-species case because coalescent times are modeled as single time points on a given branch \citep{Dutheil2009_coalhmm_smc}.

To address these limitations, \citet{Mailund2011_coalhmm_two_spc} reparameterized \texttt{CoalHMM} for a two-taxon species tree, where lineages from different species remain restricted from coalescing until the speciation event, but now time is split into discrete bins in an analogous way to \texttt{PSMC}.
This two-species model enables more accurate inference of the speciation time and effective ancestral population size \citep{Mailund2011_coalhmm_two_spc}, and has been extended to incorporate post-speciation migration between species \citep{Mailund2012_coalhmm_two_spc_mig}.

Building on both the three-species and two-species \texttt{CoalHMM} models, \texttt{TRAILS} jointly infers the topology and coalescent times of local genealogies from a three-species alignment \citep{Rivas-Gonzalez2024_trails}.
Unlike many SMC-based methods that operate on pairs of lineages (or over all possible pairs of lineages), \texttt{TRAILS} infers local genealogies by modeling both the topology (based on the four possible coalescent histories) and the timing of two coalescence events (using discrete time intervals along the species tree).
This approach enables unbiased inference of speciation times and ancestral effective population sizes for the three-species tree without requiring fossil calibrations. Moreover, \texttt{TRAILS} can reconstruct the ARG for the three species from its posterior decoding, allowing researchers to not only recover demographic parameters associated with the speciation history, but also study the underlying process of speciation itself.
By analyzing the posterior distribution of coalescence times from \texttt{PSMC}, \citep{cousins_deep_2024} showed that a diploid genome sampled from a present day human, chimpanzee, or gorilla does not fully coalesce until beyond 10Mya, and that \texttt{PSMC} is relatively robust to model violations in these ancient time periods. This indicates that coalescent-based inference can be extended much further into the past than previously thought, and in particular that a diploid primate genome contains enough information to elucidate the speciation process with its closest relatives.

\section{SMC in ARG Inference}
The SMC (and its extensions) is a full probabilistic model describing the distribution of sample genealogies.
While it is reasonably simple to \emph{simulate} a genealogy under such a model (as described in Section \ref{sec:primer}, Algorithm \ref{alg:smcprime}), the problem of \emph{inferring} or \emph{reconstructing} plausible genealogies conditional on a given sample of sequences is notoriously difficult, due to the huge search space involving both discrete (ARG topology) and continuous (branch length) components. 
Tremendous progress in ARG inference has been made in the past decade using various simplifying approximations to the SMC \citep[for recent review articles on ARGs, see][]{brandt2024promise, lewanski2024era, nielsen2024inference, Wong2024_arg}.

\citet{rasmussen2014genome} developed \texttt{ARGweaver}, a Bayesian method which outputs a sample from the posterior distribution of possible ARGs for a given genomic sample. 
\texttt{ARGweaver} relies on the idea of constructing an ARG by ``threading'' in sequences conditioned on the partial ARG of the previously added sequences, which is conceptually similar to the CSD framework \citep{Paul2011_csd_smc, Paul2010_csd, Steinrucken2013_csd_smc_structured}.
The core threading procedure reconstructs the ARG for \(n\) lineages by determining the set of branches and coalescence times for the \(n^{th}\) lineage to join the partial ARG of \(n-1\) lineages, in a manner that is consistent with both the SMC model and the observed genetic variation.
Given that the threading procedure is based on the underlying dynamics of the SMC, it is implemented as an HMM, where the hidden states represent the possible branches and coalescence times for the lineage being threaded to join the partial ARG, the transition probabilities describe the probability of branch-specific subtree prune and regraft operations under the SMC, and the emission probabilities correspond to the likelihood of the observed sequence data at the current genomic interval conditioned on the marginal genealogy, which is computed using Felsenstein's pruning algorithm \citep{Felsenstein1973_pruning, Felsenstein1981_pruning}.
\texttt{ARGweaver} relies on a Markov chain Monte Carlo (MCMC) sampler to explore the vast space of possible ARGs conditioned on the data and pre-specified model parameters (\eg, mutation rate,  recombination rate, and the demography).
Each MCMC iteration involves removing a branch (or subtree) across all marginal genealogies and then re-threading it using the HMM.
This allows for the quantification of uncertainty in both the ARG topology and estimated event times, but the HMM requires time to be discretized, and the procedure is too computationally intensive to scale beyond around 100 haploid genomes \citep{Hubisz2020_holy_grail}. To overcome the assumption that all input lineages coalesce in a panmictic ancestral population, 
\texttt{ARGweaver-D} \citep{Hubisz2020-argweaverd} was introduced as a demography-aware extension that allows explicit modeling of gene flow events between distinct input lineages. The method was used to infer that $\sim$3\% of the Neanderthal genome derives from an introgression from the early ancestors of modern humans  200kya - 300kya. More recently, \texttt{SINGER} \citep{deng2024robust} further enhanced the scalability of the threading approach by improving the efficiency of the HMM and MCMC sampler, enabling ARG inference for up to a couple of hundreds of haploid genomes (with human-like parameters). The inference from \texttt{ARGweaver} has been shown to be powerful for downstream applications such as fitting Bayesian population size history \citep{Palacios2015_gp_ne} or identifying allele frequency changes to learn about selection \citep{stern_approximate_2019, vaughn_clues2_2024}.

In contrast to these Bayesian methods that provide posterior samples, other methods which scale to much larger datasets forego the goal of principled uncertainty estimation in the ARG topology and/or event times, and instead rely on approximations to infer a single sensible ARG.
For example, \texttt{Relate} \citep{speidel2019method} and \texttt{tsinfer}/\texttt{tsdate} \citep{kelleher2019inferring,wohns2022unified} both use a modified Li-Stephens (LS) haplotype copying model \citep{Li2003_ls_model} to reconstruct local genealogies along the genome.
On the other hand, \texttt{ARG-Needle} \citep{zhang2023biobank} and \texttt{Threads} \citep{gunnarsson2024scalable} reconstruct the ARG using a threading procedure, but instead of using MCMC to sample from the posterior distribution, they use a set of ``threading instructions'' to sequentially graft a haploid genome into the partial ARG, which results in a single inferred ARG after iteratively threading each haploid genome.
While these methods do not provide estimates of uncertainty in both topology and event times like their Bayesian counterparts, they are able to scale to (hundreds of) thousands of haploid genomes.

Apart from being key to their methodology, the SMC has also been used as a null model to assess the accuracy of these tools.
Quantities calculated from reconstructed ARGs can be compared to their expectations under the SMC. 
\citet{deng2021distribution} and \citet{mckenzie2022estimating} used the SMC to derive the distribution of the genomic distance between successive local trees, and \citet{ignatieva_length_2024} derived various distributions characterizing the genomic span of clades and edges in the ARG. 
This strategy presents an alternative to breaking up reconstructed ARGs into local trees and applying tree-based metrics to measure how close reconstructed ARGs are to the simulated ground truth \citep{speidel2019method, kelleher2019inferring, yc2022evaluation}.

\section{Empirical Considerations}
Like all inference tools, estimates generated by SMC-based methods are limited by the properties and quality of the available data.
This issue is particularly pronounced when studying non-model systems with limited existing genomic resources.

\subsection{Data requirements}
SMC-based HMMs estimate the coalescence rate (inverse \Ne) at each recombination block from the set of variants observed within each block (see Section \ref{sec:primer:hmm}).
The number of linked SNPs (HMM observations) in a single recombination block improves the estimate of the genealogy at that locus (HMM hidden state). 
The total number of recombination blocks across the sequence becomes the number of independent \textit{tMRCA} estimates, and in turn, modulates the precision of the \Ne{} inference. 

\subsubsection{Genotyping resolution}
\sloppy
The properties of an organism's genome and the proportion of its bases that can be confidently genotyped can limit the number of SNPs and blocks available to analyze \citep{Nadachowska2016}.
The precision and stability of SMC-derived parameter estimates improve with the ratio of the mutation rate (\(\theta\)) to the recombination rate (\(\rho\)), scaled by the proportion of the genome that is confidently genotyped (\(p\)). 
Although SMC-based methods generally remain robust even with fragmented datasets \citep{Patton2019}, \citet{Liu2017} demonstrated that \texttt{PSMC} estimates are typically reliable when
\(\sfrac{\theta}{\rho} > 0.5\), even when applied to datasets produced by reduced representation sequencing.

Low sequencing coverage can lead to inaccuracies in genotype calls, as factors such as sequencing technology, sample quality, and filtering criteria impact the ability to confidently call genotypes. 
Population genetic tools such as \texttt{ANGSD} \citep{Korneliussen2014_angsd} have circumvented the need for strict genotype calls by utilizing genotype likelihoods.
However, SMC-based methods have yet to adopt a framework that accounts for genotype uncertainty, with \texttt{ARGweaver} and \texttt{ARGweaver-D} being two notable exceptions that can use phred-likelihoods or genotype-likelihoods as input for ARG inference.

A promising way to mitigate issues with sequencing quality is to incorporate heritable markers beyond SNPs. 
Recently, \citet{Sellinger2024} introduced an SMC-based inference method \texttt{SMCtheo} that theoretically can accommodate any marker type.
Analyzing hyper-mutable markers alongside SNPs has the potential to improve accuracy of inferred demographic histories over more recent timescales. 
For instance, \texttt{SMCtheo} has been successfully applied with Single Methylated Polymorphisms (SMPs).
\citet{Sellinger2024} used both SNP and SMP markers simultaneously to estimate recent population bottlenecks in \textit{A. thaliana}. 
However, not all hyper-mutable markers are suitable; for example, the lengths of Differentially Methylated Regions surpass the typical genomic distances between recombination events, making them incompatible with the SMC framework.

\subsubsection{Genome phasing}
Even with high-quality genotype calls, whether the genomic data is phased or unphased is a key constraint. 
Because the coalescent process models the ancestral relationships between sampled \emph{haplotypes}, rather than individuals, many SMC-based methods require phased data.
This requirement limits researchers, as statistical phasing is only possible for large sample sizes, and its alternatives require large, phased reference panels.
Meeting either condition can be challenging for systems outside of humans. 
A notable exception is \texttt{PSMC} and its variants, which only require homozygous/heterozygous calls from single diploid individual.
\texttt{MSMC2} can infer historical population sizes using unphased genomes, but its inferences are restricted to within-individual comparisons (\ie, not across pairs of genomes).
This mode results in lower resolution compared to phased datasets, and cannot be used for population separation analyses or inference of historical migration rates using \texttt{MSMC-IM}.

Nevertheless, there are a handful of SMC-based methods that circumvent the necessity of phased data,
One such approach is \texttt{MiSTI}, which provides an attractive alternative to \texttt{MSMC-IM}.
\texttt{MiSTI} only considers two diploid individuals, and bypasses the need for phasing by utilizing coalescence rates inferred by \texttt{PSMC} along with the jSFS between the two individuals.
Other SMC-based methods capable of handling unphased data include \texttt{ARGweaver} and \texttt{ARGweaver-D}.
When the input data is unphased, these methods perform ARG inference by integrating over all possible phasings during their threading procedure.

\subsection{Recency drawback}
A drawback of many SMC-based methods is their limited resolution in recent time frames.
SMC-based inference works by inferring a local coalescence rate between two sequences from the mutation density (Section \ref{sec:primer:hmm}).
When considering only two sequences, the majority of coalescence events are expected to occur in the more distant past, resulting in a dearth of information for SMC-based methods to exploit for making accurate inferences in more recent time periods \citep{Nadachowska2016, Patton2019, Santiago2020, terhorst2017robust}.
This issue is particularly concerning for conservation research, which often relies on a limited number of sequences to study the drivers of recent changes in population sizes.

While this limited resolution in recent time frames is most pronounced in SMC-based methods that make inferences from a single diploid individual (\eg, \texttt{PSMC} and its subsequent variants), methods such as \texttt{MSMC2} have better resolution for this time frame because they make inferences from multiple individuals.
However, \texttt{MSMC2}'s ability to infer recent changes in population sizes is confounded by phasing errors \citep{terhorst2017robust}.
In response to this confounding effect of phasing errors on demographic inference, \citet{terhorst2017robust} developed \texttt{SMC}$_{\text{\texttt{++}}}$, as a phase-invariant alternative. 
This strategy leverages information from the rest of the samples to inform the genealogies of the haplotype pair being analyzed. 
Specifically, it improves the reliability of \Ne{} estimates, especially for notoriously difficult recent time-scales \citep{Patton2019, terhorst2017robust}.
\texttt{SMC}$_{\text{\texttt{++}}}$ also builds on \texttt{PSMC} by accommodating larger sample sizes and applying regularization to improve estimation error (\eg, fitting smooth splines instead of assuming piecewise constant population sizes).   

Outside of the inherent theoretical challenges, SMC-based methods can also produce biased inferences of \Ne{} on more recent time scales due to a lack of \textit{a priori} demographic knowledge for the study system, or less than ideal choices of parameter values for the inference method \citep{parag_robust_2019}.
For example, \citet{leon_hilgers_avoidable_2025} reported a distinctive pattern of peaks in \Ne{} followed by a dramatic decline in recent time scales inferred by \texttt{PSMC} across many studies.
Through an extensive simulation-based and empirical sensitivity analysis, \citet{leon_hilgers_avoidable_2025} found that these biased results could be explained by unobserved population structure or solely relying on \texttt{PSMC}'s default settings for specifying the discrete time windows, which were designed for studying humans.
Mitigation strategies include changing the discretization of time to identify statistical artifacts in recent time bins and validating results with simulated data.
However, accurately inferring recent demographic changes remains a challenge for all SMC-based approaches that solely rely on a single diploid sequence \citep{Sellinger2021, leon_hilgers_avoidable_2025}. 

\subsubsection{SFS as a summary statistic for recent times}
An alternative to SMC-based methods are those that rely on the distribution of allele frequencies in a population, \ie, the site frequency spectrum (SFS).
SFS-based approaches are especially useful for recent temporal resolution, as the SFS is a more informative summary statistic for recent demographic events \citep{yang_uncertainty_2018, Patton2019}. 
SFS methods have shown enough resolution to model recent complex migration events in highly heterogeneous human populations \citep{serradell_modelling_2024}, and to trace a more complete picture of the demographic context of ancient samples \citep{kamm_efficiently_2020, sumer_earliest_2024}. 
However, their accuracy is influenced by the sample size used to construct the SFS \citep{yang_uncertainty_2018, Patton2019, terhorst_multi-locus_2015}.
SFS-based models assume independence among SNPs and do not rely on linkage disequilibrium, making them a flexible alternative when no reference genome is available, for lower-depth whole genome sequencing, and even for reduced representation sequencing data \citep{Lesturgie2022, liu_stairway_2020, excoffier_fastsimcoal2_2021}. However, the SFS suffers identifiability and sensitivity issues \citep{myers_allelic_2008, bhaskar_descartes_rule_2014, terhorst_fundamental_2015, baharian_decidability_2018, rosen_geometry_2018} which can lead to poorly supported inferences of population history \citep{deng_artifact_2025, cousins_insufficient_2025}. 

\section{Conclusions and Further Directions}
The SMC has proven to be a powerful and versatile model, enabling researchers to study questions about demographic history, natural selection, mutation and recombination rates, life history traits, structural variation, population structure, gene flow, and speciation.
As SMC-based methods tackle more complex inference tasks and accommodate less stringent data requirements, there is a pressing need to validate these methods considering demographic parameters that reflect more realistic and complex biological and ecological scenarios.
While tools are often validated using simulations, these may not fully capture the complexities of natural sequences \citep{Wang2021}.
Furthermore, as demonstrated by \citet{leon_hilgers_avoidable_2025}, users must also validate their specific implementation choices with simulations to ensure that inferences are consistent with the observed sequence data, allowing for robust interpretation of results.
Given that even modest amounts of phasing errors can bias SMC-based inferences \citep{terhorst2017robust}, it would be desirable for future methods to remain agnostic to phase-level resolution.
This would avoid potential sources of bias and enable the application of SMC-based methods to a wider range of empirical systems where phasing is unavailable or unreliable.
Extending SMC-based methods to handle low-coverage data would also be beneficial, broadening their applicability to more non-model systems, which often have fewer genomic resources, as well as to the ever-growing datasets of low-coverage ancient genomes \citep{Mallick2024_aadr}.

In addition to addressing data-level biases, model-level biases must also continue to be addressed.
Current approaches like \texttt{MSMC-IM}, \texttt{cobraa}, and \texttt{MiSTI} highlight that the assumption of panmixia is a common model violation in SMC-based inference.
However, these methods model population structure differently, and further exploration of alternative structured models is warranted.
In a similar vein, with \texttt{TRAILS} demonstrating the ability to infer the ARG for three species, extending such approaches to incorporate models of speciation with gene flow would be valuable, especially since introgression has been more recently increasingly recognized as a frequent process across the eukaryotic tree of life \citep{Dagilis2022_introgression}.
While SMC-based methods for simulating and studying structural variation have been developed \citep{peischl_sequential_2013,ignatieva_length_2024}, further work is needed to make these applicable to long-read sequencing datasets, to more fully understand the evolution of SVs of different types.

Another avenue for methodological work is to directly examine the transition matrix of SMC-based HMMs \citep{Sellinger2021}.
ARG inference methods offer a way to estimate this transition matrix directly from data, which has been explored as input for SMC-based inference \citep{strutt_joint_2023}.
In certain cases, the posterior distribution can be decoded exactly---without the use of a discrete HMM---which avoids biases introduced by the discretization of time \citep{Ki2023, schweiger_ultrafast_2023}.
As shown by \citet{cousins_structured_2025}, the transition matrix itself can be used to discern between demographic histories of panmixia versus population structure.
This approach highlights the potential for using the transition matrix as a summary statistic within an Approximate Bayesian Computation framework to test more complex competing demographic models.

Methods based on the SMC continue to emerge and prove useful for questions beyond historical effective population size.  
Accommodating alternative targets and non-standard coalescence processes has made SMC-based inference methods relevant beyond human population genetics.
Concurrently, statistical methods for reconstructing ARGs from sequencing data have turned to the SMC to efficiently generate and evaluate genealogies.
We hope to bring awareness of new SMC-derived approaches into ecology, conservation, and natural history to support their creative use in empirical studies, as well as highlight their relevance as we move into the ARG era.

\section*{Acknowledgments}
AD acknowledges support from start-up funds provided by the University of Southern California.
AN acknowledges support from the European Union's Horizon 2020 research and innovation programme under the Marie Sk\l{}odowska-Curie grant agreement No 945304 -- Cofund AI4theSciences hosted by Université PSL.
DP and ETC acknowledge support as trainees under the Brown University Predoctoral Training Program in Biological Data Science (NIH T32 GM128596).
DP also acknowledges support from the Blavatnik Family Foundation Graduate Fellowship.

\section*{Data Availability}
Code to reproduce Figure 2B has been uploaded to \url{https://gist.github.com/David-Peede/64c53ebe1f1f31d0376d33c7e9fffdd3}.
Figures 1-4 have been uploaded to \url{https://github.com/David-Peede/PDiagrams/tree/main/smc-review}.

\bibliography{smc_v_submission}
\bibliographystyle{GBE-ish}

\end{document}